%% file: main.tex
\documentclass[conference]{IEEEtran}
\IEEEoverridecommandlockouts
\usepackage[
  backend=biber,
  maxcitenames=2,
  minbibnames=3,
  mincitenames=1,
  maxbibnames=3,
  uniquelist=false,
  style=ieee,
  citestyle=numeric-comp,
  url=false,
  doi=true,
  eprint=true,
  isbn=false,
]{biblatex}
\AtBeginBibliography{\footnotesize}
\DeclareBibliographyCategory{showinbib}

\newbibmacro*{title+link}[1]{%
  \iffieldundef{doi}
    {%
      \iffieldundef{url}
        {#1}%
        {\href{\thefield{url}}{\textcolor{teal}{#1}}}%
    }
    {\href{https://doi.org/\thefield{doi}}%
       {\textcolor{teal}{#1}}}%
}

\DeclareFieldFormat
  [online,article,inbook,incollection,inproceedings,patent,thesis,unpublished]
  {title}{%
    \usebibmacro{title+link}{#1}%
  }

\renewbibmacro*{doi+eprint+url}{}
\renewbibmacro*{url+urldate}{}

\AtEveryBibitem{
	\clearfield{pages}
	\clearfield{isbn}
	\clearfield{eprint}
	\clearfield{issn}
	\clearfield{abstract}
	\clearfield{volume}
	\clearfield{number}
	\clearfield{month}
}

\usepackage{amsmath,amssymb,amsfonts}
\usepackage{algorithmic}
\usepackage{graphicx}
\usepackage{textcomp}
\usepackage{xcolor}
\usepackage{float}
\usepackage{verbatim}
\usepackage{braket}
\usepackage{xspace}
\usepackage{tikz}
\usepackage[hidelinks]{hyperref}
\usepackage{orcidlink}
\usepackage[acronym]{glossaries-extra}
\definecolor{lfdblack}{HTML}{000000}
\definecolor{lfdyellow}{HTML}{E69F00}
\definecolor{lfdgray}{HTML}{999999}
\definecolor{lfd4}{HTML}{009371}
\definecolor{lfd5}{HTML}{BEAED4}
\definecolor{lfd6}{HTML}{ED665A}
\definecolor{lfd7}{HTML}{1F78B4}

\newcommand{\eg}{\emph{e.g.}\xspace}
\newcommand{\ie}{\emph{i.e.}\xspace}
\newcommand{\etal}{\emph{et al.}\xspace}

\setabbreviationstyle[acronym]{long-short}
\glssetcategoryattribute{acronym}{nohyperfirst}{true}

\newacronym{mps}{MPS}{matrix product state}
\newacronym{mpo}{MPO}{matrix product operator}
\newacronym{dmrg}{DMRG}{density matrix renormalisation group}
\newacronym{sat}{SAT}{Boolean satisfiability}
\newacronym{nisq}{NISQ}{noisy intermediate-scale quantum}
\newacronym{pbf}{PBF}{pseudo-boolean function}
\newacronym{qubo}{QUBO}{quadratic unconstrained binary 
optimization}
\newacronym{hpc}{HPC}{high-performance computing}
\newacronym{qaoa}{QAOA}{quantum approximate optimisation algorithm}
\newacronym{pubo}{PUBO}{polynomial unconstrained optimisation}
\newacronym{lsr}{LSR}{local structure reduction}
\newacronym{sa}{SA}{simulated annealing}
\newacronym{cnf}{CNF}{conjunctive normal form}
\newacronym{gs}{GS}{ground state}
\newacronym{qc}{QC}{quantum computing}
\hypersetup{
    colorlinks=true,
    linkcolor=lfd4,
    citecolor=lfd4,
    urlcolor=lfd4
}

\begin{document}
    
\title{Towards Tensor-Network SAT-Solvers for Quantum-Classical Workflows}

\author{
    \IEEEauthorblockN{Benjamin Zec$^*$\orcidlink{0009-0009-9865-8217}, Lukas Schmidbauer$^*$\orcidlink{0009-0001-7171-0865}, Maja Franz$^*$\orcidlink{0000-0002-2801-7192}, Wolfgang Mauerer$^{*\dagger}$\orcidlink{0000-0002-9765-8313} }
    \IEEEauthorblockA{
        \textit{$^*$Technical University of Applied Sciences Regensburg, Regensburg, Germany } \\
        \textit{$^\dagger$ Siemens AG, Foundational Technology, Munich, Germany} \\
        \{benjamin.zec, lukas.schmidbauer, maja.franz, wolfgang.mauerer\}@othr.de
    }
}

\maketitle

\input{Abstract/Abstract}

\input{Introduction/Introduction}
\input{Foundations/Foundations}

\input{Methods/Methods}
\input{Results/Results}

\input{DiscussionandOutlook/DiscussionOutlook}

\input{Acknowledgment/Acknowledgment}
\input{References/References}

\end{document}

%% file: Abstract/Abstract.tex
\begin{abstract}
Integrated HPC/QC systems aim to combine classical high-performance computing with quantum processors, but cannot be reduced to mechanisms for dispatching quantum kernels. An integrated architecture must support aspects such as observability, which cannot be implemented using QPUs alone, as well as fallback execution and cost-aware decisions on whether to replace quantum tasks with classical surrogates. Such mechanisms must be approximate or benefit from problem structure to soften the inescapable exponential classical worst-case complexity. 
In this work, we study tensor-network ground-state search, as such a surrogate, for optimisation problems. This combines key quantum primitives with advanced classical simulation. It provides initial empirical indicators for surrogate selection criteria, and exposes end-to-end toolchain effects that may be missed when transformation steps are studied in isolation.
We compare a native \gls{pubo}-to-higher-order-Ising and a quadratised \gls{qubo}-to-quadratic-Ising formulation for Max-3-SAT. Both are encoded as \gls{mpo} and optimised using \gls{dmrg} approaches, with \gls{sa} as classical performance baseline. Our results show that quadratisation is not a neutral transformation step: auxiliary variables and pairwise couplings substantially degrade solution quality relative to the native higher-order representation, while \gls{sa} matches or outperforms \gls{dmrg} across all tested instances. Since the optima of \gls{sat}-derived problems are classical product states, \gls{dmrg}'s advantages don't materialise here. These findings suggest that surrogate selection in HPC/QC runtimes must be encoding- and instance-aware and provide empirical groundwork for informed decisions on fallback strategies and architecture co-design.

\end{abstract}

%% file: Introduction/Introduction.tex
\section{Introduction}

While an efficient classical simulation of future application-scale quantum computers is intrinsically impossible, numerical simulation of certain aspects nonetheless plays a crucial role for tasks such as benchmarking~\cite{arute2019quantum}, designing error-correction mechanisms~\cite{Campbell_2017}, 
sub-problems task offloading~\cite{elsharkawy2025integration,Kaya2024,schulz2023accelerating},
and especially systems-level integration~\cite{ramsauer:25:towards_qal,elsharkawy2025integration}:
The rules of quantum mechanics make it impossible to implement classically trivial properties like observability, which must be implemented based on classical (approximative) surrogates. We use a highly performant, yet adaptable simulation method in a quantum optimisation setting to study the end-to-end integration of such mechanisms into HPC/QC platforms. Tensor networks methods, in particular \gls{dmrg}, are among the most effective approaches for simulating one-dimensional quantum systems~\cite{Schollw_ck_2011}. \gls{sat} problems are classically difficult optimisation tasks. By casting \gls{sat} as an Ising problem, optimal solutions can be obtained by finding the \gls{gs} of the Hamiltonian. Different transformation paths alter key structural properties such as the number of clauses and variables~\cite{schmidbauer2024s, Gabor:2022tij}.
Efficiency and convergence of tensor networks depend on such properties, and transformation specifics influence properties of the surrogate mechanism.

We study and benchmark \gls{dmrg} across different \gls{sat}-to-Ising compilation paths and compare results against \gls{sa}, building on the analysis of Schmidbauer \etal~\cite{schmidbauer2024s,schmidbauer2025sat} to evaluate the effect of encodings, discuss the impact on \gls{hpc}-\gls{qc} hybrid workflows.

%% file: Foundations/Foundations.tex
\section{Foundations of Tensor Networks}

Our notations follows Schollwöck~\cite{Schollw_ck_2011}.
A tensor is a multilinear map that generalises matrices to arbitrary rank. Higher-rank tensors arise naturally in quantum mechanics, where the state
of an $N$-qubit system lives in a $2^N$-dimensional Hilbert space and can be represented
by a rank-$N$ tensor. Tensor networks provide a graphical and computational framework
for efficiently representing such high-dimensional tensors by decomposing them into
networks of lower-rank tensors connected by shared indices. Consider a quantum state $|\psi\rangle$ of a system with $l$ sites and $\sigma_i$ local state spaces on sites $i = 1,...,l$, each carrying local
physical dimension $d_i$. The coefficients $c_{\sigma_1 \dots \sigma_l}$ of the full
state
\(
    |\psi\rangle = \sum_{\sigma_1 \dots \sigma_l} c_{\sigma_1 \dots \sigma_l}
    \, |\sigma_1 \dots \sigma_l\rangle
    \label{eq:psi_full}
\)
can be decomposed into a product of rank-3 tensors $M_1, \dots, M_l$ , yielding the
\gls{mps} representation %
\(
    |\psi\rangle = \sum_{\sigma_1, \dots, \sigma_l}
    M_1^{\sigma_1} M_2^{\sigma_2} \ldots M_l^{\sigma_l}
    \, |\sigma_1 \sigma_2 \dots \sigma_l\rangle,
    \label{eq:MPS}
\)
where neighbouring tensors are connected by virtual bond indices of dimension $\chi$, also called
\emph{bond dimension}. This parameter limits the maximum representable entanglement, reducing the exponential memory scaling of the full state vector to $\mathcal{O}(l d \chi^2)$~\cite{Schollw_ck_2011}.
\subsection{Matrix Product Operator}

An \gls{mpo} is the tensor network representation of an
operator acting on a many-body system, directly analogous to an \gls{mps} for states. A Hamiltonian $\hat{H}$, expressed as an \gls{mpo}, reads \(
    \hat{H} = \sum_{\substack{\sigma_1 \dots \sigma_l \\ \tau_1 \dots \tau_l}}
    W_1^{\sigma_1 \tau_1} \cdots W_l^{\sigma_l \tau_l}
    \, |\sigma_1 \dots \sigma_l\rangle\langle\tau_1 \dots \tau_l|,
    \label{eq:MPO}
\)
where each local tensor $W_i$ has rank 4 with dimensions $(d_i, d_i, w_{i-1}, w_i)$ and
$w$ is the \gls{mpo} bond dimension. The physical indices $\tau_i$ and $\sigma_i$ correspond to the local $d$-dimensional input and output state spaces of the operator, respectively.

\subsection{Density Matrix Renormalization Group}
\gls{dmrg} is a variational algorithm that approximates the \gls{gs} of a Hamiltonian by iteratively optimising each tensor of an \gls{mps} to minimize $\langle\psi| \hat{H} |\psi\rangle$.
To efficiently evaluate $\langle\psi| \hat{H} |\psi\rangle$, left and right environment tensors are contracted recursively from the canonical-form \gls{mps} and stored, reducing the energy expectation to a local contraction at the site being optimized. \gls{dmrg} optimises each site tensor $M_i$ by performing iterative sweeps over the entire chain. During each sweep, the MPS tensors are updated sequentially using a sparse eigensolver (\eg, Davidson algorithm) to solve the corresponding local eigenvalue problem. We employ the two-site \gls{dmrg} variant, which jointly optimises pairs of neighbouring tensors.

\section{The Benchmark Problem, \gls{sat}}

The seminal NP-complete \gls{sat} problem serves as textbook benchmark for computational complexity. A \gls{sat} instance asks
whether there exists an assignment of Boolean variables $\{x_1, \dots, x_n\} = V$ that makes a
given formula true. Formulas are commonly expressed in \gls{cnf} \(\bigwedge_{i=1}^{m} C_i\) such that each of the \(m\) clauses $C_i$ is a disjunction of literals (variables or their negations, \eg, $C_1=x_1 \lor x_2 \lor \neg x_3$). %
In a $k$-\gls{sat} instance, each clause contains exactly $k$ literals. Max-$k$-\gls{sat} seeks to maximise the number of satisfied clauses. The experiment uses \gls{sat} formulated  as a \emph{\gls{pbf}} %
$f : \{0,1\}^n \rightarrow
\mathbb{R}$, which admits a unique multi-linear polynomial representation \(
    f(x_1, \dots, x_n) = \sum_{S \subseteq \{1,\dots,n\}} \alpha_S \prod_{j \in S} x_j,
\)
where each monomial has degree $|S|$ and coefficient $\alpha_S \in \mathbb{R}$. A \gls{pubo} problem consists of finding 
$\vec{x} \in \{0,1\}^n$ that minimises or maximises $f$ of arbitrary degree, whereas \gls{qubo} restricts $f$ to a polynomial of degree at most two. Reducing a higher-order \gls{pbf} to a quadratic one by introducing auxiliary variables $\vec{y}$ that encode the higher-order terms is called quadratisation. %

%% file: Methods/Methods.tex
\section{Experimental Setup}
\begin{figure}[t]
    \centering
    \includegraphics[width=1\columnwidth]{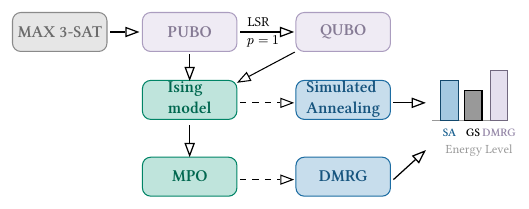}\vspace*{-1.2em}
    \caption{We benchmark two transformation paths: MAX $3$-\gls{sat} cast into \gls{pubo}, Ising and then MPO (path one) and MAX $3$-\gls{sat} cast into \gls{pubo}, quadratised to \gls{qubo}, Ising and then MPO.
    We compare \gls{sa} and \gls{dmrg} ground-state deviations for the Ising and \gls{mpo} models, respectively.
    }
    \label{fig:Pipeline}
\end{figure}

To evaluate the impact of different problem-transformation pathways, we implemented an end-to-end execution pipeline. This experimental workflow spans from the generation of uniform Max 3-\gls{sat} instances (\ie, counting satisfied clauses) to the final ground-state optimisation, as illustrated in \autoref{fig:Pipeline}. 
The experiments were conducted on a node equipped with two AMD EPYC 7662 (256 cores in total) and one TiB RAM. 

We provide an extensive \href{https://github.com/lfd/SAT_Tensor_Networks}{reproduction package} (link in pdf) that allows for extending our work. 

\subsection{SAT-Instance Generation}

We construct each clause by randomly sampling $k=3$ literals from literal 
pool $L = V \cup \{\neg x : x \in V\}$, where $V$ denotes the total set 
of variables and $\neg x$ defines the negation of $x$. This sampling 
process is repeated $m$ times to yield the total number of clauses. The 
resulting clause-to-variable ratio $\alpha = m/n$ varies across tested 
configurations. For small $n$, the number of distinct 3-literal clauses is bounded by 
$\binom{2n}{3}$, so instances with $m$ approaching this bound yields 
high clause degeneracy.
\vspace{-0.6em}
\subsection{Max 3-SAT to PUBO Transformation}
Consider a Max-$k$-\gls{sat} formula in \gls{cnf} containing \(m\) clauses $C_i$ with $k=3$ literals each. Using a standard algebraic mapping, encoding positive literals as $x_i$ and negated literals $\neg x_i$ as $(1 - x_i)$. Summing over individual clause functions yields the global objective function for the \gls{pubo} problem \(f_{\text{SAT}}(\vec{x}) = \sum_{i=1}^{m} f_{C_i}(\vec{x})\). Since $f_{C_i}(\vec{x}) = 1$ if $C_i$ is satisfied and $0$ otherwise, $s = f_{\text{SAT}}(\vec{x})$ evaluates to the number of satisfied clauses. Casting Max-3-\gls{sat} as maximisation problem where perfect satisfiability yields $s = m$. Expanding 3-\gls{sat} polynomials yields non-linear monomials up to degree 3 \cite{schmidbauer2024s}.
\vspace{-0.5em}
\subsection{PUBO to QUBO}

To transform a higher-order \gls{pubo} $f_p$ into a \gls{qubo} $f_q$, we use the \gls{lsr} algorithm~\cite{schmidbauer2024s}, an iterative quadratisation technique governed by a free parameter $p \in [0,1]$ that manages the trade-off between the number of introduced auxiliary variables and the connectivity of the resulting interaction graph. A higher value of $p$ minimises the number of auxiliary variables at the expense of a denser degree-2 interaction matrix in $f_q$.  As each additional introduced variable doubles the search space, we fix $p = 1$ in our automated pipeline used in our experiments.
\vspace{-0.5em}
\subsection{Ising to MPO Transformation}
To evaluate the optimisation performance via \gls{dmrg}, \gls{pubo} and \gls{qubo} formulations are mapped onto their respective Ising representations and encoded as an \gls{mpo}. For the quadratised \gls{qubo} pipeline, the substitution yields a standard quadratic Ising Hamiltonian consisting of (at maximum) two-body Pauli-$Z$ coupling terms. 
Conversely, the Hamiltonian in the \gls{pubo} pipeline inherits higher-order terms as multi-body spin interactions.
Both Hamiltonians are explicitly constructed as \gls{mpo}s according to \autoref{eq:MPO}. 
Auxiliary variables required to reduce  higher-order terms to quadratic forms increase the effective problem dimension and introduce additional coupling terms not present in the native formulation~\cite{biamonte2008nonperturbative}.
\begin{figure*}[t]
    \input{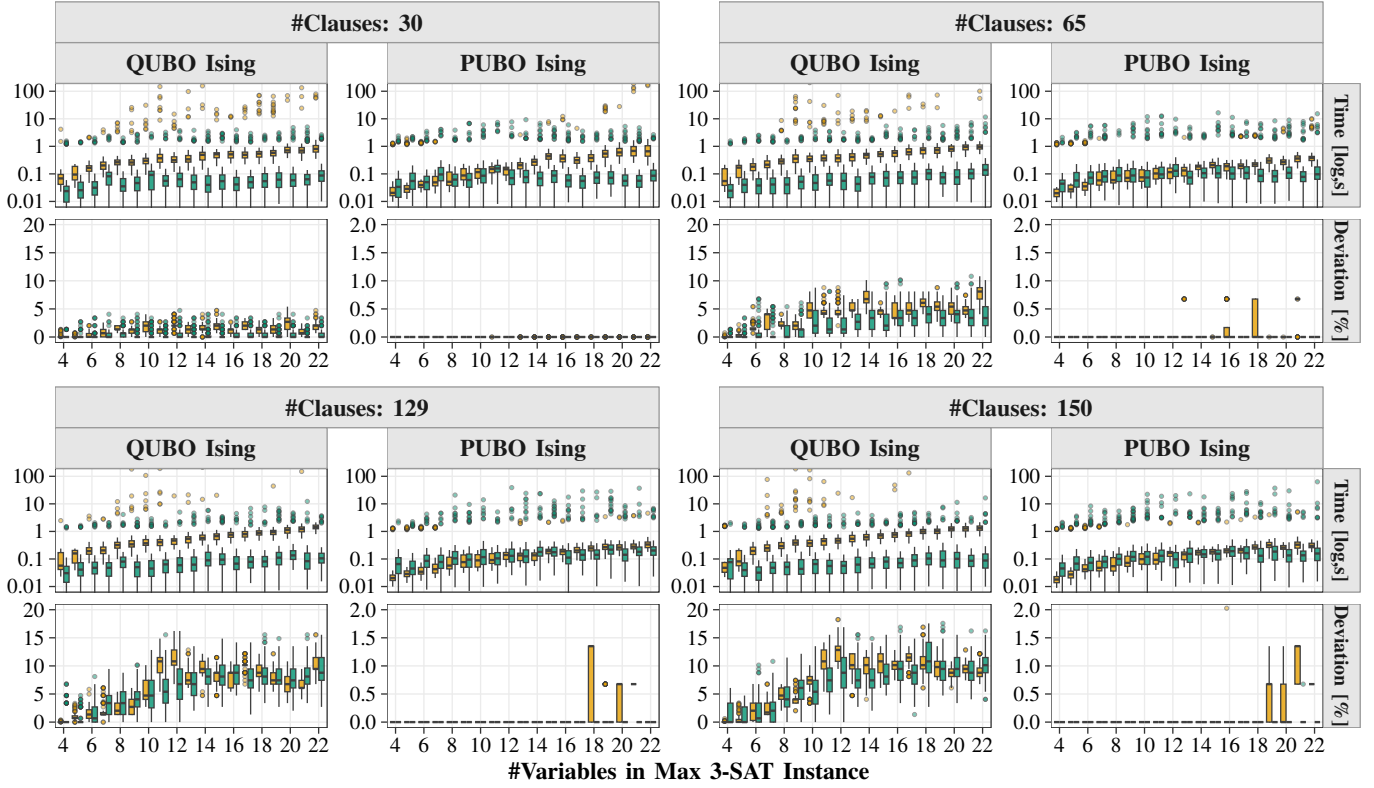}\vspace*{-2.2em}
    \caption{
      Optimisation performance of \gls{sa} (green) and \gls{dmrg} (yellow),
      grouped by number of SAT variables~$n$ and clause count~$m$.
      Upper panels show runtime per instance: \gls{sa} is consistently faster and
      \gls{dmrg} scales with $n$ in both encodings, where outliers reflect instances
      with increased bond dimension~$\chi$, driving up contraction cost.
      Lower panels show deviation from the ground state: the \gls{pubo} encoding
      keeps deviations near zero throughout, while \gls{qubo}
      deviations grow monotonically with $n$ and $m$ and outliers 
      reflect individual instances trapped in local minima due to insufficient
      bond dimension~$\chi$.
    }
    \label{fig:Results}
\end{figure*}
\subsection{Solver Configuration and Hyperparameter Benchmarking}
\vspace{-0.2em}
To evaluate optimisation performance and eliminate initialisation bias, each Max 3-SAT instance is evaluated over 100 independent trials using randomised hyperparameters in both solvers. For the \gls{dmrg} framework, we used a vanilla two-site setup, as it provides a conservative lower bound on what a more specialised tensor-network approach might achieve. Each trial starts from a random \gls{mps} ansatz, randomly varying the number of sweeps between 5 and 13, drawing the truncation error cutoff $\epsilon \in \{10^{-9}, 10^{-11}, 10^{-13}, 10^{-14}\}$, and sampling a terminal maximum bond dimension $m_{\text{max}} \in \{20, 40, 50, 100, 200, 300\}$. To ensure cross-encoding comparability, the ground-state energy of each \gls{qubo} Ising Hamiltonian is corrected by adding the Ising constant. The \gls{sa} baseline executes a localised spin-flip kernel with a stochastical schedule for each trial. The initial temperature $T_0$ is drawn from $[5.0, 150.0]$, the final stop temperature $T_{\text{min}} \in \{10^{-3},10^{-4},10^{-5},10^{-6}\}$, the cooling rate $\alpha \in [0.985, 0.999]$, and the Monte Carlo steps per temperature step from $[30, 500]$. Upon completion, the results of all 100 runs per instance are aggregated and compared against the exact global minimum extracted from the \gls{pubo} energy landscape and its computational time per instance. 

%% file: Results/Results.tex
\section{Results}
\autoref{fig:Results} summarises the optimization performance achieved by \gls{dmrg} and \gls{sa}.

\paragraph{Deviation from ground state}
As the number of variables $n$ in \gls{sat} increases, Ising formulations diverge.
In the Quadratic Ising (\ie, \gls{qubo}) regime, the median deviation for both solvers grows monotonically with $n$, rising from below $1\%$ at small instance sizes to deviations exceeding $10\%$ to $15\%$ for $n \ge 12$ in the facets where $m \in  \{129,150\}$. This performance follows instance-dependent solution quality drops, observed for \gls{qubo}-encoded  $3$-SAT on quantum annealing platforms~\cite{gabor2019assessingsolutionquality3sat}. The interquartile range and high-deviation outliers both widen with $n$, indicating that larger \gls{qubo} instances are not only solved less accurately on average but also less reliably across independent trials.
In contrast, for both solvers in the higher-order Ising case, the corresponding boxes remain close to zero deviation across the entire tested range of variables (note the chosen scale in $[0,2]$ for this encoding).
Localised deviations and minor outliers for \gls{dmrg} within the higher-order pipeline only begin to manifest at the absolute highest complexity bounds where $m \ge 129$ and $n \ge 18$.
The gap between the two encodings widens with instance size, since the \gls{qubo} pipeline's auxiliary-variable overhead and pairwise-interaction density both scale with $m$.
Comparing the two solvers directly within the quadratic encoding, \gls{sa} matches or slightly outperforms \gls{dmrg} for the majority of instances, and this performance gap widens as $n$ and $m$ increase, particularly past $n = 12$ where the upper whiskers in the \gls{dmrg} case extend significantly higher than those in \gls{sa}.
\paragraph{Runtime} 
Runtime measurements show a similar asymmetry: \gls{sa} remains consistently faster than \gls{dmrg} across all configurations. \gls{dmrg}'s runtime increases steadily with $n$ in both encodings, exceeding $10s$ for moderate sizes and approaching $120s$ for the densest $m=150$ instances. Notably, this runtime growth is not confined to the quadratic encoding as in the higher-order panels, where deviation remains close to zero across nearly the entire tested range, \gls{dmrg}'s runtime nevertheless scales with $n$ in essentially the same manner as in the quadratic case.

This indicates that \gls{dmrg}'s cost is driven by \gls{mpo} contraction and variational sweeping as $n$ grows. These combined metrics demonstrate that \gls{dmrg}'s defining strength, namely efficiently representing many-body entanglement at bounded bond dimension, provides no advantage in this experiment, where \glspl{gs} are inherently classical product states. Taken together, these results suggest that for \gls{sat}-derived combinatorial problems, the \gls{pubo} representation outperforms the \gls{qubo} representation in both runtime and ground state solution quality, since the auxiliary variables and additional coupling terms introduced during quadratisation appear to affect \gls{dmrg} performance.

%% file: DiscussionandOutlook/DiscussionOutlook.tex
\section{Discussion and Outlook}
Our principal technical result is that quadratisation is not a transparent compilation step. For the studied Max-3-SAT instances, the native \gls{pubo} encoding yields markedly better \gls{gs} search results than the corresponding \gls{qubo} formulation, while \gls{sa} matches or outperforms \gls{dmrg}. Tensor-network methods thus offer no immediate advantage for this problem class, whose optima are classical product states. Spending additional classical computational resources may reduce the implementation overhead of \gls{dmrg}, as tensor contractions and eigensolver steps admit distributed-memory and multi-GPU acceleration~\cite{Levy2020}. This would
not, however, remove the encoding-induced difficulties. This distinction matters directly for integrated \gls{hpc}/\gls{qc} systems. Such systems require more than the dispatch of quantum kernels: they must decide when a quantum task should be replaced by a classical surrogate, and must expose sufficient information to support debugging and scheduling~\cite{delgado2025defining}. Our results show that this decision must be encoding- and instance-aware. Performance models proposed in Refs~\cite{thelen:25:qce25, Thelen_2024} provide a natural mechanism: measurements such as those reported here can serve as baseline data for predicting solution quality and cost, and hence for selecting between \gls{sa}, \gls{dmrg}, and quantum execution. Within such a runtime, \gls{dmrg} is best regarded not as a universally superior solver, but as a bounded classical surrogate.
Where tractable, it can be a key component in implementing full (state) observability
and provide an independently reproducible
reference for debugging quantum subroutines~\cite{rovara2025framework}, which is intrinsically unachievable with a pure quantum approach (including under \gls{nisq} imperfections that complicate direct
diagnosis~\cite{greiwe2023effects}). It can also support approximate fallback execution when quantum resources are unavailable.

Future work should turn this characterisation into explicit policies for surrogate selection, cost-aware offloading, and observability hooks in \gls{hpc}-\gls{qc} runtimes. The broader lesson is that compiler and runtime design must account for the complete transformation path: representation choices can dominate solver behaviour and should be treated as first-class decisions.

%% file: Acknowledgment/Acknowledgment.tex
\begin{small}
  \noindent\textbf{Acknowledgements} This work was supported by the German Federal Ministry of Research, Technology and Space (BMFTR), funding program ‘Research Program Quantum Systems’, grant number 13N17387, the High-Tech Agenda of the Free State of Bavaria, and the German Research Foundation, grant MA 9739/1-1.
\end{small}

%% file: References/References.tex
\printbibliography